# RANDOM RELAY SELECTION BASED HEURISTIC OPTIMIZATION MODEL FOR THE SCHEDULING AND EFFECTIVE RESOURCE ALLOCATION IN THE COGNITIVE RADIO NETWORK

Aravindkumaran.S1 *, Dr.Saraswady.D2

1Research scholar, Department of Electronics and communication engineering, Puducherry Technological University, Puducherry, India.
2Professor, Department of Electronics and communication engineering, Puducherry Technological University, Puducherry, India.

## ABSTRACT

*Cognitive Radio Network (CRN) provides effective capabilities for resource allocation with the valuable spectrum resources in the network. It provides the effective allocation of resources to the unlicensed users or Secondary Users (SUs) to access the spectrum those are unused by the licensed users or Primary Users (Pus). This paper develops an Optimal Relay Selection scheme with the spectrum-sharing scheme in CRN. The proposed Cross-Layer Spider Swarm Shiftingis implemented in CRN for the optimal relay selection with Spider Swarm Optimization (SSO). The shortest path is estimated with the data shifting model for the data transmission path in the CRN. This study examines a cognitive relay network (CRN) with interference restrictions imposed by a mobile end user (MU). Half-duplex communication is used in the proposed system model between a single primary user (PU) and a single secondary user (SU). Between the SU source and SU destination, an amplify and forward (AF) relaying mechanism is also used. While other nodes (SU Source, SU relays, and PU) are supposed to be immobile in this scenario, the mobile end user (SU destination) is assumed to travel at high vehicle speeds. The suggested method achieves variety by placing a selection combiner at the SU destination and dynamically selecting the optimal relay for transmission based on the greatest signal-to-noise (SNR) ratio. The performance of the proposed Cross-Layer Spider Swarm Shifting model is compared with the Spectrum Sharing Optimization with QoS Guarantee (SSO-QG). The comparative analysis expressed that the proposed Cross-Layer Spider Swarm Shifting model delay is reduced by 15% compared with SSO-QG. Additionally, the proposed Cross-Layer Spider Swarm Shiftingexhibits the improved network performance of ~25% higher throughput compared with SSO-QG.*

## 1. INTRODUCTION

Cognitive Radio Networks (CRN) perform learning and communication with the neighboring environment to recognize and access the space in the spectrum to reduce the frequency of occurrence [1]. CR is defined as the "intelligent radio network technology for the variation in the transmitter parameters those vary dynamically based on operating environment for the self-mobility behaviour to achieve environmental awareness and transfer in parameters". In the case of signal processing, CR is considered the smart wireless communication system those acquires knowledge from the environment with the use of artificial intelligence technology [2]. CR comprises the statistical properties computed based on the signal variation in wireless systems based on the real-time change in operational parameters [3]. Like radio technology, CR uses cognitive intelligence for the real time environment for the effective utilization of available spectrum through modification in the factor and decision-making process. The CR comprises of the precondition of the available spectrum effectively with the categorization of primary (PUs) (authorized users) and secondary users (SU) (unauthorized users). To estimate the openness in the spectrum idle spectrum is sensed based on the periodic estimation of user identity [4]. The SU utilizes the available frequency band without affecting on the PUs communication.

In CRN routing and scheduling is considered as the challenging task that computes the diversity of the channel availability and data rates [5]. The conventional scheduling and routing schemes in CRN exhibit the some limitations with the weighted max-min fair scheduler those impact on the window size, central scheduler and exchange of information from the central scheduler between each user, issues related to fairness, minimized throughput, increase in maintenance cost and so on [6]. Due to reduced channel availability, the throughput of the network is minimized. The CRN routing process is efficient with minimized time and energy utilization with the appropriate resource scheduling schemes [7]. Conventionally, CRN resource allocation techniques increase the consumption of power and reduce interference [8]. The unused spectrum or holes are replaced with the integration of PUs with the SUs. The number of relay nodes or SUs increases the performance of the secondary network with reduced performance characteristics [9]. The spectrum sharing technique is implemented for the effective distribution of spectrum between secondary users with appropriate usage of cost. The selection of effective channels through spectrum sharing relies on the secondary users in CRN to compute the detection of effective spectrum [10].

Research Contribution and motivation:

Cognitive radio improves SE through controlling spectrum use. Additionally, as compared to the conventional one-way relaying, two-way relaying boosts SE. Together with CD, these two methods greatly improve the SE. To increase the security of the secondary user network, we take into account single and multiple relay selection strategies. Analysis reveals that while employing a single relay, secondary transmitter(ST) and secondary destination (SD) merely selects the best secondary relay for secondary transmission. However, if we choose numerous secondary relays, the secondary signal is sent to the destination through multiple secondary relays. We calculate the transmission intercept probability and outrage probability for both relay selection systems using spectrum sensing. The relay selection strategies are mathematically analysed using RT, and direct transmission is offered for comparison. According to the aforementioned situation, the reliability of spectrum sensing is raised or the likelihood of a false alarm is decreased by utilising a filter and forward relay transmission. As a result, both relay selection techniques' RTs have increased.

## 2. RELATED WORKS

Cross-layer design comprises of the optimized design model with the concept of waterfall in Open System Interconnection (OSI) reference model provides the interconnection between the protocol layers to provides the interdependences between layers and adaptability for the information exchange between layers [11]. The cross layer interactions provides the appropriate increase in the efficiency of the network and Quality of Service (QoS). In network wireless technology cross-layer design is considered as the effective network model for the transmission of information through physical medium those are significant over the time [12]. The information exchange between the layers is computed based on the optimized throughput of the network state. The examination expressed that energy consumption with cross-layer model is less effective [13]. The implementation of the cross-layer design is considered as the effective method to increase the performance of the mobile ad hoc network. The performance of the network is computed based on the consideration of different parameters such as reliability, delay, throughput, and energy efficiency [14]. The effective spectrum and sharing between users are performed with the physical layer and the spectrum decision process is performed in the network layer. The implementation of the spectrum mobility with the appropriate channel switching is performed with the physical layer-based spectrum sensing model for the allocation of resources in link layer and routing in network layer [15]. For the purpose of managing interference in underlay CRNs, many signal processing approaches, including spread spectrum, power regulation, and beamforming, have been investigated [16]. Power control controls interference in the power domain, whereas spread spectrum manages interference in the code domain. Beamforming uses the spatial degrees of freedom (DoF) that numerous antennas give to guide secondary signals in specific directions and keep main users free from interference. Beamforming is more appealing in practise compared to the other two approaches since it is good at reducing interference. Given its potential, beamforming has been investigated in underlay CRNs to pursue a variety of goals, including enhancing the security against eavesdroppers [19], maximising the data rate of secondary users [18], and improving the energy efficiency of secondary transmissions [17].

Research Gap:

The networking and productivity of CRN are constrained owing to the degree of allowed interference to either the PUs or even among the SUs themselves, which is another significant issue that makes RA problems in CRN particularly difficult. The most severe impediment to maximising resourcefulness and usefulness in CRN is likely the restriction on SUs' gearbox caused by the degree of permitted interference to PUs. For the reasons listed above, it is essential to conduct in-depth research on the foundation and guiding principles for implementing/adapting the existing methodologies for RA in other wireless communications to CRN. Such investigations will assist to characterise and examine its workability as well as determine its suitability of application or purpose. This literature assessment demonstrates that much work has already been done in this area, but that much more work is still needed to fill up the research gaps. Therefore, knowing optimisation is essential to comprehending RA challenges in CRN and devising solutions. Essentially, optimisation may be investigated and used as a crucial strategy for resolving RA issues in CRN. As a well-developed analytical technique addressing a wide range of issues, optimisation is widely employed in a variety of scientific disciplines, including operations research, engineering, corporate finance, and economics. In optimisation, there is typically a goal that has to be accomplished, such as maximising or minimising one or more entities. This goal is always reflected in the objective function. Then, when working to accomplish the goal, there are some restricting factors that must be taken into account. Constraints must be broken during problem-solving; otherwise, any solutions that may be found are null and void. The choice variables are generally often the last element of an optimisation issue.

In CRN, MAC layer comprises of sharing of spectrum, selection of relay, sharing of resources and scheduling is performed. In other hand, network layer performs the routing with the effective management of spectrum through network and MAC layer. Hence, in this paper presented an effective routing and scheduling scheme for the selection of relay to improve cross-layer design is presented. The cross layer design comprises of MAC layer rules forwarded to the network layer in the sender side. The Spider Swarm based optimization model is implemented for the optimal path selection and routing. The packets within the network are presented with dynamic packet shifting model to increase the throughput and reduce delay. The proposed CLSSS model comprises of the cross-layer design for the computation of path in the network. The optimal paths are estimated with the SSO with the data shifting process.

This paper is organized as : section II discuss existing technique based on CRN spectrum sharing. Section III provides the existing works on the resource allocation in CRN. Section IV provides the architecture of the proposed CLSSS model and simulation results are presented in Section V and overall conclusion is presented in Section VI.

## 3. MATERIALS AND METHODS

### 3.1. System Architecture: RSPSO

The proposed Cross-Layer Spider Swarm Shifting (CLSSS) comprises of the 4 modules such as CRN, decision manager, optimization module and rule-based modules is illustrated in the figure 1. Within CRN modules it comprises of two modules those are relay selection and spectrum allocation modules. The SUs designed to support the PUs in the network and Sus are designed to offer effective relay selection with a selection of efficient network relay. The relay in the network are computed based on the effective relay selection algorithm for the decision-making process. The process comprises of spatio temporal constraints, co-ordination communication model for the rule based decision management process for the Optimization process using SSO model. All data process is controlled by the decision manager with the proposed CLSSS. The CLSSS perform sharing of spectrum through dynamic allocation of resources between PUs and SUs. The CLSSS model derive the optimal decision-making process with dynamic packet shifting integrated with the Spider Swarm Optimization module formed rules.

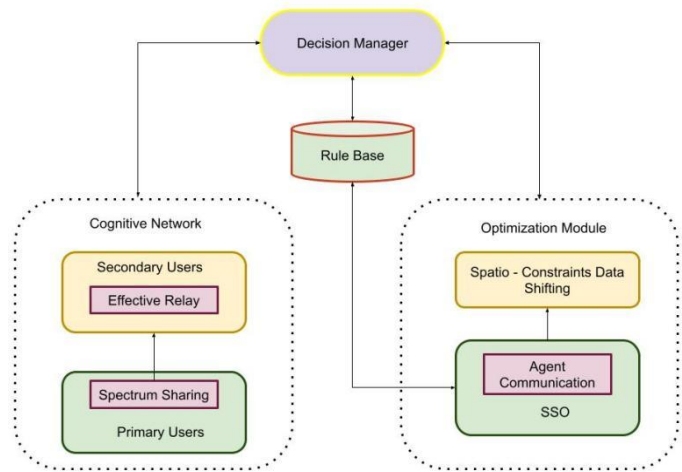

Figure 1: Overall Architecture of CLSSS

The dynamic shifting is implemented based on the framed rules applied over the spider warm optimization module, spatio-temporal management and communication-based on agent.

The decision manager stores those rules and manipulate the process for the dynamic packet switching. Moreover, with the spatio-temporal analysis integrated with SSO perform module amplification and random relay forwarding and selection for the secondary network. The CLSS improves the QoS based on the prearranged values computed for the spectrum for the decision manager request for the allocation of resources to the PUs. The primary network comprises of the unused spectrum those integrates the Secondary network with spider swarm optimization. The network throughput is increased by the presented algorithm model.

## 3.2. System Modelling

The path for data transmission between PUs and SUs estimated with CLSSS model is presented in figure 2. With the implementation of secondary network cooperative relay is estimated for the spectrum efficiency estimation and data rate computation. In CRN available spectrum band channels are allocated between the primary and secondary users. The frequency bands or holes for PUs are applied over the relay nodes those dataset are computed with spider swarm optimization. The developed CLSSS model compute the communication in relay nodes with the estimation of data rate. Secondary network comprises of the one source node denoted as A and the destination node in the network is denoted as B. The nodes have the M relay denoted as $\{R_1, R_2, \ldots \ldots, R_M\}$. The CLSSS compute chain gain with the estimation of link between source to relay node and destination is denoted as $CG_{Ai}$ and $CG_{RBi}$ respectively. Those relay nodes are measured with the CLSSS model performance based on consideration of the two cooperation model such as Spatio-temporal factor and Amlify and Forward (AF). With the increase in node relay the network throughput decreases. To overcome the issue spatio-temporal packet shifting is implemented to achieve optimal data packet forwarding with application of spatio-temporal constraints.

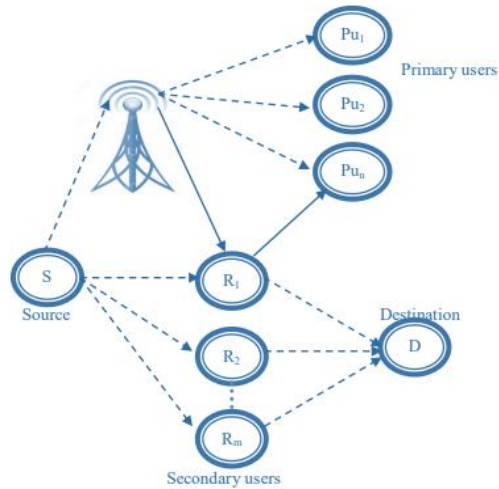

Figure 2: Data Transmission between PUs and SUs

## 3.3. Optimal Relay Selection

With the proposed CLSSS model optimal relay is selected through spatiotemporal relay selection constraints model. It comprises of the ideal secondary user for the signal transmission for the relay selection in the signal. The relay selection algorithm provides the effective relay selection scheme with the spatio-temporal constraints based on the Signal to Noise Ratio (SNR) with the selection of relay in CRN. Based on the computed SNR values the frames are estimated for the relay to transmit the signal transmission between transmission to the destination node. The node capable to act as the relay is represented as in equation (1) to equation (3)

$$Z_{SRi} = \sqrt{w_s} CH_{SRi} y_s + n_{SRi}, \quad i = 1,2, \ldots \ldots, M \quad (1)$$
$$Z_{R,D} = \alpha CH_{R,D} z_{SRi} + n_{R,D} \quad (2)$$
$$Z_{R,D} = \alpha \sqrt{W_s} CH_{R,D} CH_{SRi} y_s + \alpha CH_{R,D} n_{SRi} + n_{R,D} \quad (3)$$

International Journal of Computer Networks & Communications (IJCNC) Vol.15, No.6, November 2023In the above equation the transmitted signal resource node is defined as RS which involved in construction of signal to every receiver. The link in the destination node comprises of the received relay noise with the source relay comprises of the additive Gaussian noise. The pertaining transmission power with Source A subjected to interference of PUs. The relay amplification factor is indicated as α and the relay node power and amplification factor to satisfy the succeeding condition presented in equation (4)

$$\alpha = \sqrt{\frac{W_s}{(|CH_{SRi}|^2 W_s + N_o)}} \qquad (4)$$

In above equation (4), the average noise power for each channel in secondary network is presented as $N_o$. Based on the consideration of equation (1) to equation (3) the SNR of the destination node in trails are computed as in equation (5) to equation (7)

$$\eta_{SRi} = \frac{W_s|CH_{SRi}|^2}{(N_o)} \quad (5)$$
$$\eta_{RD} = \frac{W_s|CH_{RiD}|^2}{(N_o)} \quad (6)$$
$$\eta_{R,D}^{\Delta F} = \frac{W_s|CH_{SRi}|^2|CH_{RiD}|^2\alpha^2}{\alpha^2|CH_{RiD}|^2 + N_o + N_o} \quad (7)$$

The equivalent SNR $\eta_{R,D}^{\Delta F}$ is computed using equation (6) and (7), the resultant equation is presented in (8)

$$\eta_{R,D}^{\Delta F} = \frac{\eta_{SRi}.\eta_{RD}}{\eta_{SRi} + \eta_{RD} + 1} \qquad (8)$$

The signal to noise ratio is computed for the relay nodes with the varying coefficient of channel fading. Consequently, the prerequisite values are computed based on the SNR threshold valuesthose lies between A and B and the set of relay node constructed is denoted as $U(R)$. The receiver SNR value for the ith SNR for the relay node and threshold is presented in equation (9)

$$\eta_{SRi} = \frac{W_s|CH_{SRi}|^2}{(N_o) \geq \eta_{th}} \qquad for\ i = 1,2,\ldots,M \qquad (9)$$

In case true node relay it is included in the set or eliminated. Upon the computation of similar process for the all-nodes new sets are computed using the equation (10)

$$U(R) = \{(R_i|\eta_{SRi} \geq \eta_{th})\}, \qquad i = 1,2,\ldots,M \quad (10)$$

The process of relay selection is performed for the different channel sender and receiver nodes. The packet delivery ratio of the node set are presented in equation (11)

$$V(R) = \{(R_i|\max(\eta_{R,D}))\}, \qquad i = 1,2,\ldots,M \quad (11)$$

| Algorithm 1:Resource Allocation with CLSSS |
|---|
| 1. Compute the received signal ith relay node noise with spatio-temporal computation.<br>2. Estimate the node solution based on the threshold of the relay nodes<br>3. Compute the link between the nodes towards the destination for the channel data transmission with highest noise ratio.<br>4. Select the relay node in the network<br>5. Compute the relay link between sender to receiver using the equation (10)<br>6. Perform the spatitemporal constraints in the channel link |

154

7.   Identify the best relay link in the network

With the consideration of above equation (10) and (11) the best effective possible relay link are computed between sources to destination node. The best relay link estimated is presented in equation (12)

$$R_{best} = \{V(R) \cap U(R)\} \qquad (12)$$

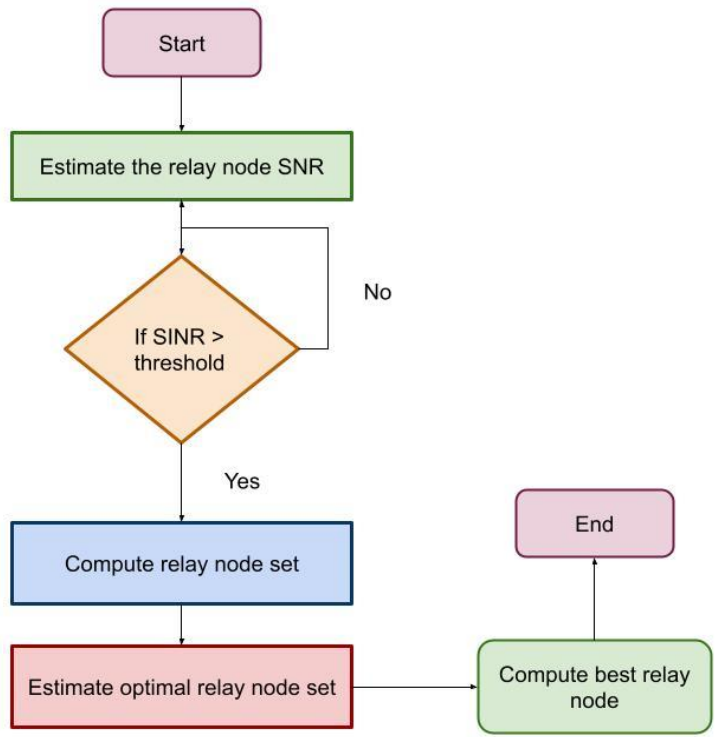

Figure 3: Process in CLSSS

The proposed CLSSS model process is presented in figure 3 with the estimation of relay node in the candidate node, optimal and best.

### 3.4. Spectrum Sharing with Spider Swarm Optimization

The proposed CLSSS model comprises of the secondary links count of N or the primary relay link denoted as S for the mathematical modelling. With the primary channel data were transmitted through the network relay links. Based on the similar characteristics estimation relay link is estimated for the primary and relay channel. The primary link in the network provides the assigned exclusive primary channel for the relay link within the network. The channel interference and capacity is computed with the utilization of the utilized spectrum as stated in equation (13)

$$\gamma_j = \frac{W/sld(j)^m}{\sum_{l-\emptyset}\frac{W}{rsd(l,j)^m} + \frac{W}{psd(i,j)^m}}, \quad 1 \leq j \leq M \qquad (13)$$

As in figure 4 the Signal to Noise Ratio (SINR) is computed for the computation of primary and secondary links. The secondary receiver SINR is represented as $\gamma_j$, the transmitter power value is denoted as $W$ and attenuation factor is denoted as $m$. The distance between the computed signal to noise ratio is denoted as $sld(j)^m$, $rsd(l,j)^m$ and $psd(i,j)^m$.

The ith primary receiver signal to noise ratio is presented in equation (14)

$$\gamma_j = \frac{W/pld(i)^m}{\sum_{l-\emptyset} \frac{W}{spd(l,i)^m}}, \quad 1 \leq j \leq M \quad (14)$$

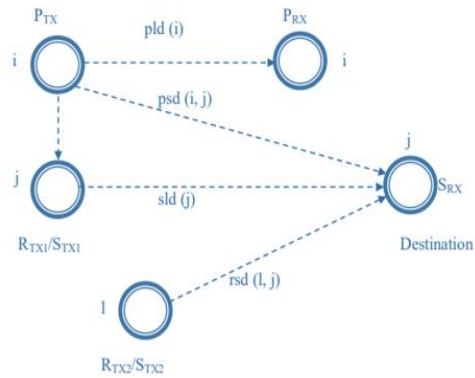

Figure 4: Calculation of SINR

Based on the subsequent constraints spectrum sharing is performed based on the condition presented in equation (15) and equation (16)

$$\gamma_i \geq \gamma_L, \quad 1 \leq i \leq N \quad (15)$$
$$\gamma_j \geq \gamma_L, \quad 1 \leq j \leq M \quad (16)$$

The primary link and relay link channel capacity is denoted as $C_i$ and $C_j$ as represented in equation (17) and (18)

$$C_i = BW \log_2(1 + \gamma_i) \quad (17)$$
$$C_j = BW \log_2(1 + \gamma_j) \quad (18)$$

In above equation (17) and (18) primary channel bandwidth is represented as $BW$ and the system general format is presented in equation (19)

$$Max \sum_{j=1}^{M} c_j y_j + \sum_{i=1}^{N} c_i \quad (19)$$

Where, $c_i > 0, c_j > 0$, which can be simplified as in equation (20)

$$y_j \in \{0,1\}, 1 \leq j \leq M \quad (20)$$

Based on the equation (18) applying the constrains the optimal path within the network is estimated. The proposed CLSSS model uses the SSO integrated with the temporal constraints.

Based on the developed model the particle velocity and position are computed and updated as in equation (21) and equation (22)

$$v_{id}(t+1) = \omega V_{id}(t) + c_1 r_1(p_{best}(t) - y_{id}(t)) + c_2 r_2(g_{best}(t) - y_{id}(t)) \quad (21)$$

$$y_{id}(t+1) = y_{id}(t) + v_{id}(t+1) \quad (22)$$

In the above equation (21) and (22) the inertia weight are represented as $\omega$ the cognitive coefficient are denoted as $c_1$ and social coefficient are denoted as $c_2$ the random values are presented as $r_1$ and $r_2$ between the range of 0 and 1. $p_{best}$ old best position of node and present & global best position of node is represented as $g_{best}$. The matrices for the particles are computed based on the dimensions S ×M with consideration of matrices. The population size is represented as $S$ and relay links of the node is presented as $N$. The primary and secondary links in channel are computed as p' best and y'. The relationship between the primary and secondary channel are presented in figure 5.

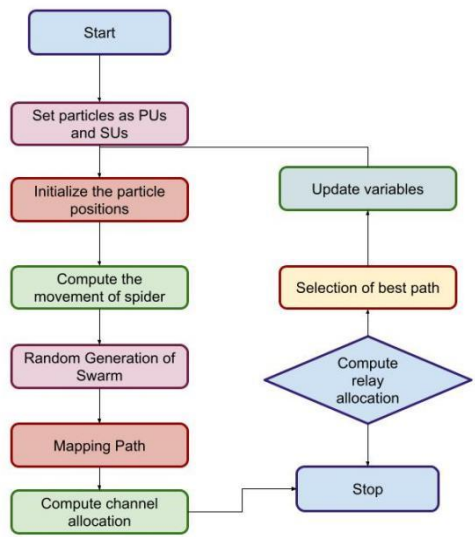

Figure 5: Flowchart of CLSSS

## 4. RESULTS AND DISCUSSION

The proposed CLSSS model is implemented in the network simulator NS2. The simulation is performed for the mobile nodes in the region coverage of 1000 m × 1000 m with the standard distance of 1 Km for the total simulation time of 100 seconds. The node mobility for the movement is considered in the range of 10 m/sec to 50 m/sec. Every node in the network exhibits the equal transmission range of 250 meters. The node count is varies between 50 – 100 nodes. The node transmission power is stated as 0.660W and receiving power of nodes are defined as 0.395W with the initial energy level of 100J. The parameters incorporated in the simulation model for the relay selection and packet shifting is presented in table 1.

Table 1: Simulation Setting of CLSSS

| Parameter | Value |
| --- | --- |
| Number of Nodes | 100 |

| Initial Energy (J) | 100 |
|---|---|
| Communication Area | $1000 \times 1000$ m |
| MAC layer | 802.11 |
| Time for Simulation (s) | 100 |
| Antenna | Omni Antenna |
| Range of Transmission(m) | 250 |
| Size of Packets (KB) | 512 |
| Power for Transmission (W) | 0.660 |
| Power of Reception (W) | 0.395 |

The secondary network incorporated node for the relay selection with the consideration of 10 relay to establish the link of 5 with the primary and secondary users. The links in relay varies from 2 -`10 with the increment of 2 in the spectrum channel for the relay selection through primary links with the optimization using spider swarm. Through the consideration of the data rate input is provided for the demanding channel for the primary links.

*Performance Based on Data Rate*

The performance is evaluated with the CLSSS model based on the consideration of average data rate. With the consideration of data rate relay link number are computed with the proposed CLSSS model 10 relay links are considered. For increase in data rate number of requesting link also increased. The relay links are computed based on the primary links controlled with the value of γ. The primary channel are assigned with 6 channel for the multimedia data transmission with the secondary network.

Table 2: Average Data Rate

| Relay Links | Data Rate (kbps) | | | | |
|---|---|---|---|---|---|
| | $\gamma = 6$dB | $\gamma = 8$dB | $\gamma = 10$dB | $\gamma = 12$dB | $\gamma = 14$dB |
| 2 | 100 | 500 | 1000 | 1500 | 2100 |
| 4 | 200 | 700 | 1100 | 1700 | 2300 |
| 6 | 350 | 900 | 1300 | 1900 | 2700 |
| 8 | 400 | 1100 | 1600 | 2100 | 3100 |
| 10 | 500 | 1200 | 1900 | 2200 | 3300 |

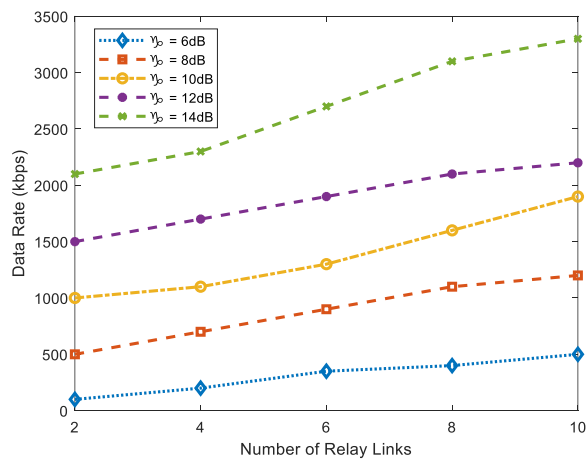

Figure 6: Average Data Rate

International Journal of Computer Networks & Communications (IJCNC) Vol.15, No.6, November 2023

The figure 6 and table-2 illustrated the number of channel increases relay link of data rate . For channel 2 the user receives the data rate of 2.8Mbps for the increase channel to access higher data rate. The data rate received by the channel 6 is measured as 6.8Mbps. The Table 3 and figure 7 provides the examination of data rare for the varying number of channel. The conducted communication traffic for the multimedia data is computed for the 15 – 35 channels. The figure 7 illustrate that the increase in data rate increases the channel and reaches the maximal channel value of 35 with the higher data rate of 10 bits/second for the channel 35. Therefore, it is observed that number of channels effectively process and transmit the multimedia data.

Table 3: Comparison of Channel Vs Data Rate

| Number of Channels/nodes | Date Rate (Mbps) | Delay (ms) |
|---|---|---|
| 20 | 3 | 7.1 |
| 40 | 4.1 | 7.6 |
| 60 | 6 | 8.2 |
| 80 | 8 | 8.4 |
| 100 | 11.3 | 8.8 |

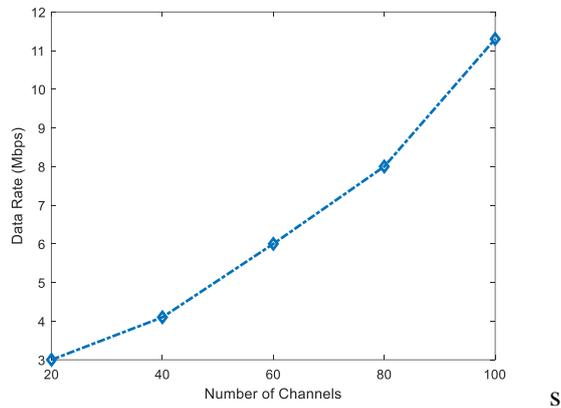

Figure 7: Comparison of Data Rate

*Performance Based on Nodes*

The performance of proposed CLSSS metrices are evaluated based on the varying number of node. The performance of CLSSS model is compared with the existing Spectrum Sharing Optimization model to provided guaranteed QoS.

Table 4: Comparison of Delay

| Number of nodes | Delay (ms) | |
| | SSO-QG | CLSSS |
|---|---|---|
| 20 | 15 | 7.1 |
| 40 | 16.3 | 7.6 |
| 60 | 18 | 8.2 |
| 80 | 16.6 | 8.4 |
| 100 | 12.2 | 8.8 |

International Journal of Computer Networks & Communications (IJCNC) Vol.15, No.6, November 2023

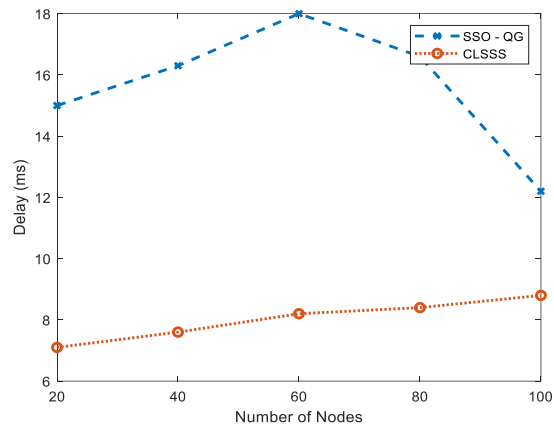

Figure 8: Comparison of Delay

The figure 8 and Table 4 illustrates the analysis of delay with the proposed CLSSS the performance of scheme is minimized the delay value of 15% compared with SSO-QG. The process of data shifting is performed for the data forwarding with the minimizes the error rate and increases the packet forwarding capacity effectively. Additionally, the CLSSS model increases the spectrum sharing capability with the reduced transmission delay. The standard energy model is utilized for the energy consumption. The figure 9and Table 5 provides the representation of energy consumption for the CLSSS model than the existing techniques.

Table 5: Comparison of Energy Consumption

| Number of nodes | Energy Consumption (J) | |
|---|---|---|
| | SSO-QG | CLSSS |
| 20 | 3.3 | 1.2 |
| 40 | 4.7 | 1.9 |
| 60 | 5.8 | 2.2 |
| 80 | 5.6 | 2.4 |
| 100 | 5.7 | 2.7 |

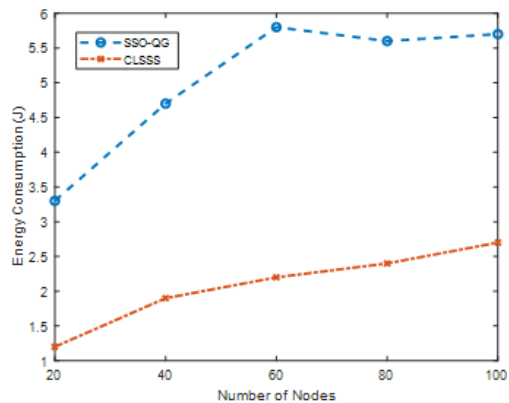

Figure 9: Comparison of Energy Consumption

International Journal of Computer Networks & Communications (IJCNC) Vol.15, No.6, November 2023

The figure 9 provides the energy consumption of the proposed CLSSS model is 13% minimal compared with the existing SSO-QG model. As the proposed CLSSS model comprises of the intelligent agents for effective communication with the SSO model. The figure 10 illustrates the comparative examination of throughput and delivery ration for the varying number of nodes.

Table 6: Comparison of Metrices

| Number of Nodes | Throughput (kbps) | | Packet Delivery Ratio | | Overhead | |
|---|---|---|---|---|---|---|
| | SSO-QG | CLSSS | SSO-QG | CLSSS | SSO-QG | CLSSS |
| 20 | 2100 | 6400 | 0.6 | 0.88 | 17 | 7 |
| 40 | 4200 | 6700 | 0.67 | 0.93 | 22 | 12 |
| 60 | 4200 | 10000 | 0.78 | 0.96 | 27 | 15 |
| 80 | 4100 | 12000 | 0.84 | 0.97 | 26 | 16 |
| 100 | 6000 | 15000 | 0.9 | 0.99 | 29 | 17 |

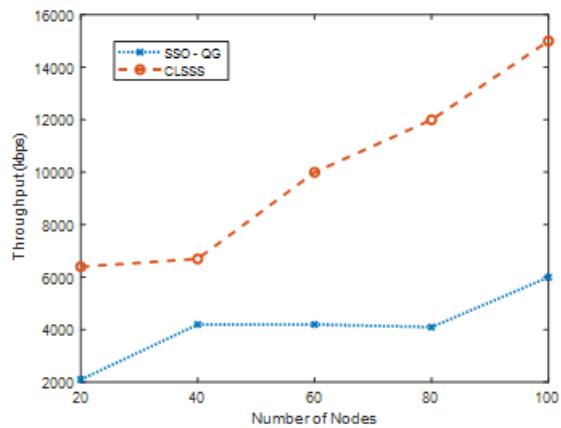

Figure 10: Comparison of Throughput

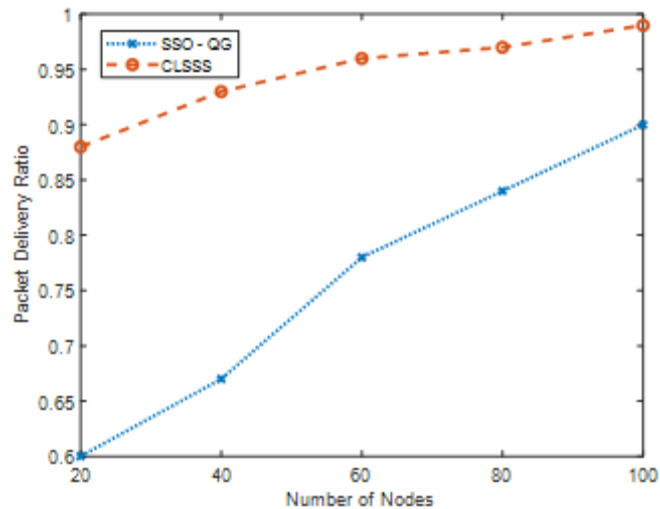

Figure 11: Comparison of packet Delivery Ratio

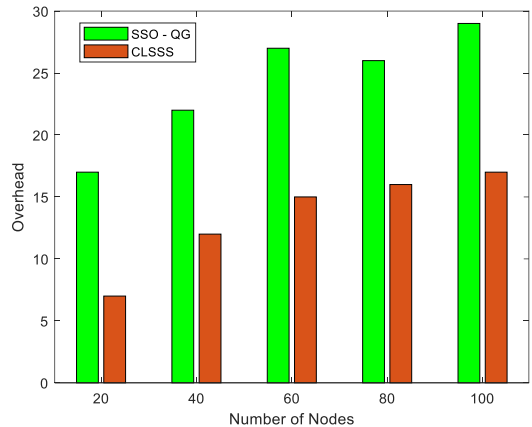

Figure 12: Comparison of Overhead

Through analysis of figure 10 and Table 6 illustrated that network throughput increases by 25% with the CLSSS compared with the SSO-QG. With the proposed CLSSS model SSO is utilized for the network estimation based on global position. The global position is evaluated based on near relay link for channel selection in primary link. The simulation analysis expressed that proposed CLSSS model increases the throughput. The comparative analysis is performed for the packet delivery ratio to the existing technique.

In figure 12 stated that packet delivery ratio of CLSSS model is increases with 10% compared with existing SSO-QG model. With the CLSSS model SSO identifies the optimal route with the shortest distance. In the figure 11 it is observed that overhead rate is minimized with the proposed CLSSS model by 10% compared with existing SSO-QG. The reduction is overhead is reduced by the relay node performance operations.

Research limitation:

We take into account an underlying CRN with two active primary users and two active secondary users. The main users set up time-division duplex (TDD) mode bidirectional communications. In the main network, traffic moves steadily and consistently in both directions. In order to use the same spectrum for their own communications, the secondary users wish to do so. To do this, the secondary transmitter uses beamforming to pre-cancel the interference it generates for the primary receiver, while the secondary receiver uses IC to detect the signal. Simply put, the secondary users bear the whole cost of cancelling out cross-network interference, while the prime users are completely unaware of their data transmissions.

## 5. CONCLUSION

CRN comprises of the heterogeneous wireless architecture for the dynamic spectrum access to achieve the effective utilization of higher bandwidth. With effective resource allocation CRN offers the valuable spectrum utilization of the network resources. For the suggested system model, the signal-to-noise ratio and outage probability are examined under various scenarios in order to explore the impact of numerous key agents that affect the system's performance. The OPs are assessed specifically under various numbers of SU relays, various numbers of the fading severity parameter of channels, various numbers of PU locations, and various speeds of the mobile end user. The simulation results are also given in order to validate the analytical results. Overall, the results demonstrated that the performance of the system may be enhanced by using a cooperative communication strategy using cognitive radio networks. Additionally, a variety of variables, including channel fading, the number of SU relays, the fading severity parameter, the PU location, and the speed of the mobile end user, might influence how well cooperative spectrum sharing networks operate with mobile end users. These elements are seen to be essential for service providers to consider while designing their networks. These crucial elements must be considered in the design and implementation of a functional mobile network. The performance is comparatively examined with SSO-QG stated that the proposed CLSSS scheme achieves the ~15% reduced delay. Through analysis concluded that CLSSS model effectively improves the overall performance with increases throughput of ~25% then the conventional techniques. The examination confirmed that proposed CLSSS model performance is effective for relay selection and resource allocation in CRN.

**CONFLICTS OF INTEREST**

The authors declare no conflict of interest.

## A<small>UHTORS</small>

**Aravindkumaran S** received his B.Tech degree in Electronics and communication engineering from Rajiv Gandhi College of Engineering and Technology, Puducherry, India, in 2015 and his M.Tech degree in Electronics and Communication Engineering from Sri Manakula vinayagar Engineering College, Puducherry, India, in 2017. He is now working towards his Ph.D in the Department of Electronics and Communication Engineering, Puducherry Technological University, Puducherry, India. His research interests are in the areas like Cognitive radio networks, Cross-layer routing.

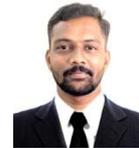

**D.Saraswady** received her B.Tech and M.Tech degrees from Puducherry Technological University Erstwhile Pondicherry Engineering College, Pondicherry University, Puducherry, India, in 1993 and 1996, respectively, and her Ph.D degree from Anna University, Tamilnadu, India, in 2006. She is currently serving as a professor in the Department of Electronics and Communication Engineering, Puducherry Technological University, Puducherry, India. Her research interest includes image processing, cognitive radio networks, wireless ad hoc networks, and enhancement of MIMO systems.

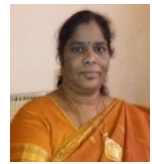